\documentclass[12pt]{article}
\pdfoutput=1
\usepackage{jheppub}
\usepackage[utf8]{inputenc}
\usepackage{amsfonts,amsmath}

\usepackage{ifthen}
\usepackage{enumerate}
\usepackage{amsmath}
\usepackage{amssymb}
\usepackage[mathscr]{eucal}
\usepackage[errorshow]{tracefnt}
\usepackage{amsmath}
\usepackage{slashed}
\usepackage{amssymb}
\usepackage{array}
\usepackage{amsthm}
\usepackage{eucal}
\usepackage{epsf}
\usepackage{epsfig}
\usepackage{psfrag}
\usepackage{multirow}
\usepackage{wasysym} 
\usepackage{tikz-cd} 
\usepgfmodule{shapes}
\usetikzlibrary{backgrounds, arrows,calc,shapes,decorations.pathreplacing, decorations.markings, automata,positioning}

\usetikzlibrary{arrows,calc}
\usepackage{relsize}

\newcommand{\be}{\begin{equation}}
\newcommand{\ee}{\end{equation}}
\newcommand{\bea}{\begin{eqnarray}}
\newcommand{\eea}{\end{eqnarray}}
\newcommand{\ba}{\begin{array}}
\newcommand{\ea}{\end{array}}
\newcommand{\bpic}{\begin{tikzpicture}}
\newcommand{\epic}{\end{tikzpicture}}

\newcommand\uot{U(1)_{\text{top}}}
\renewcommand\Im{{\mathrm{Im}}}
\renewcommand\Re{{\mathrm{Re}}}

\newcommand{\M}{{\mathfrak M}}

\newcommand{\Llra}{\Longleftrightarrow}
\newcommand{\ra}{\rightarrow}

\newcommand{\lra}{\leftrightarrow}

\newcommand\pbp{\bar{\psi}\psi}

\newcommand\phit{\tilde \phi}
\newcommand\psit{\tilde \psi}

\newcommand{\CC}{{\mathcal C}}

\newcommand{\CL}{{\mathcal L}}

\newcommand{\CT}{{\mathcal T}}

\newcommand{\cC}{\mathcal{C}}

\newcommand{\cL}{\mathcal{L}}

\newcommand{\cN}{\mathcal{N}}
\newcommand{\cO}{\mathcal{O}}

\newcommand{\bC}{\mathbb{C}}

\newcommand{\bP}{\mathbb{P}}

\newcommand{\bR}{\mathbb{R}}

\newcommand{\bZ}{\mathbb{Z}}

\def\a{\alpha}

\def\e{\varepsilon}

\def\l{\lambda} 
\def\s{\sigma}

\renewcommand{\r}{{\rho}}

\title{QED's in $2{+}1$ dimensions: complex fixed points and dualities}

\preprint{SISSA  53/2018/FISI}

\author[1,2]{Sergio Benvenuti,}
\author[1]{Hrachya Khachatryan}

\affiliation[1]{International School of Advanced Studies (SISSA), Via Bonomea 265, 34136 Trieste, Italy}
\affiliation[2]{INFN, Sezione di Trieste, Via Valerio 2, 34127 Trieste, Italy}
\emailAdd{benve79@gmail.com, hrachya.khachatryan@sissa.it}

\abstract{We consider Quantum Electrodynamics with an even number $N_f$ of bosonic or fermionic flavors, allowing for interactions respecting at least $U(N_f/2)^2$ global symmetry. Both in the bosonic and in the fermionic case, we find four interacting fixed points: two with  $U(N_f/2)^2$ symmetry, two with $U(N_f)$ symmetry. 

Large $N_f$ arguments suggest that, lowering $N_f$, all these fixed points merge pairwise and become complex CFT's. In the bosonic QED's the merging happens around $N_f\sim 9{-}11$ and does not break the global symmetry. In the fermionic QED's the merging happens around $N_f\sim3{-}7$ and breaks $U(N_f)$  to $U(N_f/2)^2$.

When $N_f=2$, we show that all four bosonic fixed points are one-to-one dual to the fermionic fixed points. The merging pattern suggested at large $N_f$ is consistent with the four $N_f=2$ boson $\lra$ fermion dualities, providing support to the validity of the scenario.

}

\begin{document}

\maketitle

\section{Introduction and summary}

Quantum Electro Dynamics (QED) in $2+1$ dimensions, with either fermionic or bosonic flavors, is a paradigmatic example of a Quantum Field Theory with a strongly coupled infrared behaviour. Both in the bosonic and in the fermionic case, if the number of flavors $N_f$ is large enough, the Renormalization Group (RG) flows to a unitary interacting Conformal Field Theory (CFT). For small $N_f$, however, other possibilities remain open.  One option is that lowering $N_f$ the RG fixed point becomes "complex": the real RG flow slows down ("walking") close to the complex fixed point, and the phase transition is "weakly first order" instead of second order, see \cite{Gorbenko:2018ncu} for a modern perspective. Starting from a ultraviolet unitary gauge theory, the mechanism is that \emph{two} real fixed points, varying continuously some parameter like space-time dimension or $N_f$, annihilate into each other and become a \emph{pair} of complex conjugate fixed points.

The case of QED$_3$ with $N_f{=}2$, two bosonic or two fermionic flavors, is particularly interesting: it describes the N\`eel --- Valence Bond Solid quantum phase transition \cite{QCP1, QCP2}, moreover non-trivial boson $\lra$ fermion dualities are expected to hold \cite{Karch:2016sxi, Wang:2017txt}. Among other things, the dualities imply symmetry enhancements to $O(4)$ or $SO(5)$ at the fixed point, depending on the model. The dualities \cite{Karch:2016sxi, Wang:2017txt} are part of recent interesting progress in $3d$ dualities, see for instance  \cite{Aharony:2011jz, Giombi:2011kc, Aharony:2012nh, Son:2015xqa, Aharony:2015mjs, Seiberg:2016gmd, Karch:2016aux, Metlitski:2016dht, Hsin:2016blu, Aharony:2016jvv, Benini:2017dus, Komargodski:2017keh, Benini:2017aed, Jensen:2017bjo, Gomis:2017ixy, Bashmakov:2018wts, Benini:2018umh, Choi:2018ohn, Choi:2018tuh}, and \cite{Senthil:2018cru} for a review.

In this paper we consider QED's with an even number $N_f$ of bosonic or fermionic flavors. We allow for quartic interactions respecting $U(N_f/2)^2$ global symmetry\footnote{We are more precise about the global structure of the symmetry group in eqs. \ref{exactsymmb1}, \ref{exactsymmb2}, \ref{exactsymmf1}, \ref{exactsymmf2}.}. In both cases we argue that there are four interacting fixed points, two with $U(N_f/2)^2$ global symmetry, two with $U(N_f)$ global symmetry. 

Bosonic QED with $N_f/2$ flavors $\phi_i$ plus  $N_f/2$ flavors $\phit_i$, and $U(N_f/2)^2$ global symmetry, has four fixed points, which we denote as
\begin{itemize}
\item{\bf{bQED}} (\emph{tricritical}), $U(N_f)$ global symmetry. Both the mass term and the quartic scalar interactions are tuned to zero, so the potential vanishes: $V(\phi_i, \phit_j)=0$.
\item{\bf{bQED$_+$}} (\emph{Abelian Higgs, or $\bC\bP^{N_f-1}$, model}), $U(N_f)$ symmetry. $V{\sim}(\sum_i |\phi_i|^2{+}|\phit_i|^2)^2$.
\item{\bf{ep-bQED}} (\emph{easy plane QED}), $U(N_f/2)^2$  symmetry. $V{\sim}(\sum_i |\phi_i|^2)^2{+}(\sum_j |\phit_j|^2)^2$.
\item{\bf{bQED$_-$}}. $U(N_f/2)^2$  symmetry. $V{\sim}(\sum_i |\phi_i|^2{-}|\phit_i|^2)^2$.
\end{itemize}
The fermionic QED's have $N_f/2$ flavors $\psi_i$ plus  $N_f/2$ flavors $\psit_i$. We introduce one or two real scalars, interacting with the fermions via cubic Yukawa couplings, such models are called QED-Gross-Neveu or QED-Nambu-Jona-Lasinio.\footnote{In the literature there are two different models called QED-Gross-Neveu: one with $3d$ global symmetry $U(N_f)$ (which we name QED-GN$_+$) and one with $3d$ global symmetry $U(N_f/2)^2$ (which we name QED-GN$_-$). For us each $\psi, \psit$ is a complex two-component  $3d$ fermion. See also footnote \ref{Fclari}.} Allowing for $U(N_f/2)^2$ global symmetry gives four fixed points, 
which we denote as
\begin{itemize}
\item{\bf{fQED}} (\emph{standard}), $U(N_f)$ global symmetry. The non-gauge interacting part of the Lagrangian, $\CL_{int}$, vanishes.
\item{\bf{QED-GN$_+$}} (\emph{QED-Gross-Neveu}), $U(N_f)$ symmetry. $\CL_{int} = \r_+ \sum_i (\bar{\psi}_i \psi_i + \bar{\psit}_i \psit_i)$.
\item{\bf{QED-NJL}} (gauged \emph{Nambu-Jona-Lasinio}), $U(N_f/2)^2$  symmetry. There are two real scalars: $\CL_{int} = \sum_\pm \r_\pm ( \sum_i \bar{\psi}_i \psi_i \pm  \bar{\psit}_i \psit_i)$.
\item{\bf{QED-GN$_-$}}. $U(N_f/2)^2$  symmetry. $\CL_{int} = \r_- \sum_i (\bar{\psi}_i \psi_i - \bar{\psit}_i \psit_i)$.
\end{itemize}

In section \ref{SEC:fixedpoints} we discuss the fixed points and report the the scaling dimensions of quadratic and quartic scalar operators, computed in the large-$N_f$ expansion in \cite{Benve:2018A}. 

For $N_f$ large enough the fixed points are real CFT's, but what can we say about the smallest possible $N_f$, namely $N_f{=}2$?

Considering all the different models together allows for a useful unified perspective. Dualities can help us. If $N_f{=}2$, the four bQED's are dual to the four fQED's, in the following fashion (\bpic   \draw[<->, ultra thick, orange] (0,0) -- (1.5,0);\epic stands for "dual to")
\be\label{4dualities}\bpic[scale=0.8]
  \draw [fill, red] (0,0) circle [radius=0.1];
  \draw [fill, red] (0,4) circle [radius=0.1];
  \draw [fill, red] (1.6,1.35) circle [radius=0.1];
  \draw [fill, red] (-2,2.9) circle [radius=0.1];
  \node at (0.7, 4.4) {ep-bQED$, U(1)^2$};
  \node at (-1.7,2.5) {bQED$_-, U(1)^2$};
  \node at (1.9,1.7) {bQED$_+, U(2)$};
  \node at (0.6,-0.4) {bQED$, U(2)$};
  
  \draw [fill, red] (9,0) circle [radius=0.1];
  \draw [fill, red] (9,4) circle [radius=0.1];
  \draw [fill, red] (10.6,1.35) circle [radius=0.1];
  \draw [fill, red] (7,2.9) circle [radius=0.1];
  \node at (9.7, 4.4) {fQED$, U(2)$};
  \node at (+7.3,2.5) {QED-GN$_-, U(1)^2$};
  \node at (10.9,1.7) {QED-GN$_+, U(2)$};
  \node at (9.6,-0.4) {QED-NJL$, U(1)^2$};
  
  \draw[<->, ultra thick, orange] (0.5,0) -- (8.5,0);
  \draw[<->, ultra thick, orange] (0.5,4) -- (8.5,4);
  \draw[<->, ultra thick, orange] (2.1,1.35) -- (10.1,1.35);
  \draw[<->, ultra thick, orange] (-1.5,2.9) -- (6.5,2.9);
  
  \epic\ee
The first and third dualities were discovered in \cite{Karch:2016sxi, Wang:2017txt}. We obtain the second and fourth dualities (which are new) in section \ref{SEC:dualities}. The dualities do not tell us if the fixed points are real or complex, but suggest that all $8$ fixed points share a similar fate at small $N_f$, and restrict the possible scenarios.

Numerical simulations in $N_f{=}2$ fQED, bQED$_+$ and ep-bQED suggest second order or weakly first order transitions with certain critical exponents \cite{LSK, Kaul:2011dqx, Karthik:2016ppr, DEmidio:2016wwg, Zhang:2018bfc}. However the numerical bootstrap \cite{Nakayama:2016jhq, DSD, Poland:2018epd, IliesiuTALK} shows that there are no $3d$ unitary CFT's with those critical exponents and $O(4)$/$SO(5)$ symmetry.

Lowering $N_f$ as a continuos variable, 
it has long been suspected that fermionic QED dynamically develops quartic interactions that break the global symmetry $U(N_f) \ra U(N_f/2)^2$ \cite{Pisarski:1984dj, Kaveh:2004qa, DiPietro:2015taa, Giombi:2015haa, DiPietro:2017kcd, Herbut:2016ide, Gusynin:2016som, Kotikov:2019rww}. On the bosonic side, \cite{MarchRussell:1992ei, Nogueira:2013oza, Nahum:2015jya, Nahum:2015vka} proposed that the $\bC\bP^{N_f-1}$ model merges with tricritical QED and become a pair of complex CFT's.

Keeping track of the $N_f=2$ dualities \ref{4dualities}, we can propose a scenario which is consistent with all the above observations/proposals: all the $8$ fixed points merge pairwise at some $N_f^*>2$ (the four merging points have different $N_f^*$), below which the fixed points are complex CFT's.

On the bosonic side, the fixed point {\bf ep-bQED} merges with {\bf bQED$_-$}, while the fixed point {\bf bQED$_+$} merges with {\bf bQED}, without breaking the global $U(N_f)$ symmetry. On the fermionic side,  {\bf fQED} merges with {\bf QED-GN$_-$}, while {\bf QED-GN$_+$} merges with {\bf QED-NJL}, breaking the global symmetry $U(N_f) \ra U(N_f/2)^2$. In a cartoon, we draw the path of the fixed points as an analytically continued function of $N_f$ (continuous line for real fixed points at $N_f{>}N_f^*$,  dashed line for complex fixed points at $N_f{<}N_f^*$, arrows go in the direction of decreasing $N_f$), and we see that the merging pattern is consistent with the four $N_f=2$ dualities:
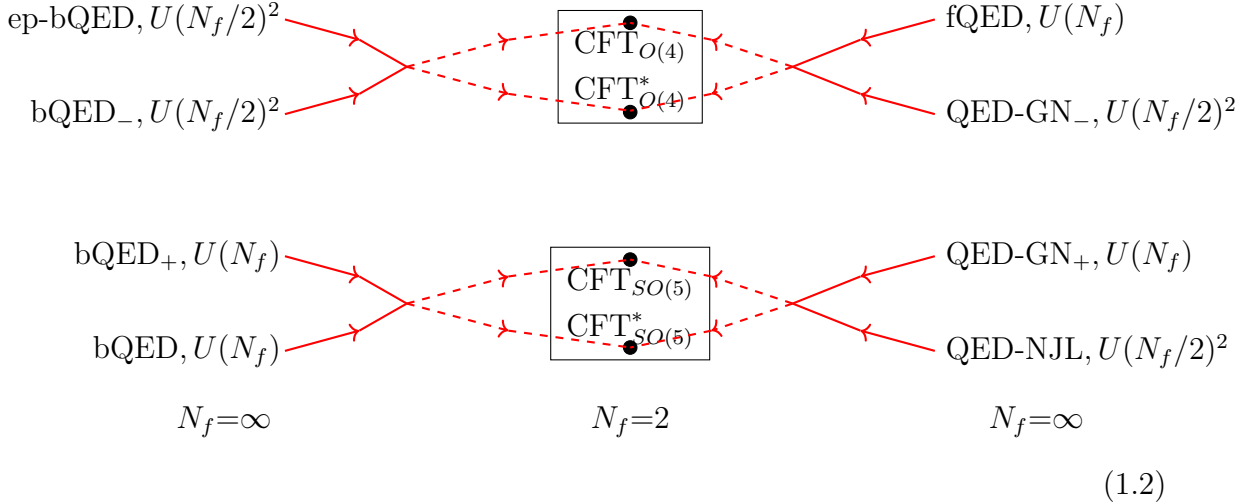
\be\label{merging}\bpic[scale=0.9]
  \draw [fill, black] (7,4.65) circle [radius=0.1];
  \draw [fill, black] (7,3.33) circle [radius=0.1];
  \draw [fill, black] (7,1.15) circle [radius=0.1];
  \draw [fill, black] (7,-0.15) circle [radius=0.1];

  \node[left] at (2, 4.7) {ep-bQED$, U(N_f/2)^2$};
  \node[left] at (2,3.3) {bQED$_-, U(N_f/2)^2$};
  \node[left] at (2,1.2) {bQED$_+, U(N_f)$};
  \node[left] at (2,-0.2) {bQED$, U(N_f)$};
  
  \node[right] at (11.5, 4.7) {fQED$, U(N_f)$};
  \node[right] at (11.5,3.3) {QED-GN$_-, U(N_f/2)^2$};
  \node[right] at (11.5,1.2) {QED-GN$_+, U(N_f)$};
  \node[right] at (11.5,-0.2) {QED-NJL$, U(N_f/2)^2$};

  \node at (1,-1.25) {$N_f{=}\infty$};
  \node at (7,-1.25) {$N_f{=}2$};
  \node at (13,-1.25) {$N_f{=}\infty$};

\node at (7,0.5) [rectangle,draw] {$\ba{c} \text{CFT}_{SO(5)}  \\ \text{CFT}^*_{SO(5)}\ea$};
\node at (7,4) [rectangle,draw] {$\ba{c} \text{CFT}_{O(4)}  \\ \text{CFT}^*_{O(4)}\ea$};

  \draw[->, thick, red] (1.9,4.7-3.5) -- (3,4.4-3.5);   \draw[-, thick, red] (3,4.4-3.5) -- (3.7,4-3.5);
  \draw[->, thick, red] (1.9,3.3-3.5) -- (3,3.6-3.5); \draw[-, thick, red] (3,3.6-3.5) -- (3.7,4-3.5);
  \draw[->, thick, red, dashed] (3.7,4-3.5) -- (5.2,4.4-3.5);   \draw[-, thick, red, dashed] (5.2,4.4-3.5) -- (7,4.65-3.5);
  \draw[->, thick, red, dashed] (3.7,4-3.5) -- (5.2,3.6-3.5); \draw[-, thick, red, dashed] (5.2,3.6-3.5) -- (7,3.35-3.5);
  
  \draw[->, thick, red] (1.9,4.7) -- (3,4.4);   \draw[-, thick, red] (3,4.4) -- (3.7,4);
  \draw[->, thick, red] (1.9,3.3) -- (3,3.6); \draw[-, thick, red] (3,3.6) -- (3.7,4);
  \draw[->, thick, red, dashed] (3.7,4) -- (5.2,4.4);   \draw[-, thick, red, dashed] (5.2,4.4) -- (7,4.65);
  \draw[->, thick, red, dashed] (3.7,4) -- (5.2,3.6); \draw[-, thick, red, dashed] (5.2,3.6) -- (7,3.35);

  \draw[->, thick, red] (11.5,4.7) -- (10.4,4.4);   \draw[-, thick, red] (10.4,4.4) -- (9.4,4);
  \draw[->, thick, red] (11.5,3.3) -- (10.4,3.6); \draw[-, thick, red] (10.4,3.6) -- (9.4,4);
  \draw[->, thick, red, dashed] (9.4,4) -- (8.2,4.4);   \draw[-, thick, red, dashed] (8.2,4.4) -- (7,4.65);
  \draw[->, thick, red, dashed] (9.4,4) -- (8.2,3.6); \draw[-, thick, red, dashed] (8.2,3.6) -- (7,3.35);

  \draw[->, thick, red] (11.5,4.7-3.5) -- (10.4,4.4-3.5);   \draw[-, thick, red] (10.4,4.4-3.5) -- (9.4,4-3.5);
  \draw[->, thick, red] (11.5,3.3-3.5) -- (10.4,3.6-3.5); \draw[-, thick, red] (10.4,3.6-3.5) -- (9.4,4-3.5);
  \draw[->, thick, red, dashed] (9.4,4-3.5) -- (8.2,4.4-3.5);   \draw[-, thick, red, dashed] (8.2,4.4-3.5) -- (7,4.65-3.5);
  \draw[->, thick, red, dashed] (9.4,4-3.5) -- (8.2,3.6-3.5); \draw[-, thick, red, dashed] (8.2,3.6-3.5) -- (7,3.35-3.5);
  \epic\ee
\paragraph{Estimation of merging points from large $N_f$.} Let us explain the rationale behind the merging scenario from the large-$N_f$ perspective. Let us consider the scaling dimensions of the quadratic operators in tricritical bQED and in fQED, at $1^{st}$ order in $1/N_f$:
\be\label{scalingbase}
\left\{\ba{rcl}
\Delta[\phi^*\phi_{adjoint}] &=& 1 - \frac{64}{3 \pi^2 N_f} \\
\Delta[\phi^*\phi_{singlet}] &=& 1 + \frac{128}{3 \pi^2 N_f}\ea \right.
\qquad \qquad
\left\{\ba{rcl}
\Delta[\bar{\psi}\psi_{adjoint}] &=& 2 - \frac{64}{3 \pi^2 N_f} \\
\Delta[\bar{\psi}\psi_{singlet}] &=& 2 + \frac{128}{3 \pi^2 N_f}\ea \right.
\ee

Decreasing $N_f$ continuously, in bQED the singlet operator approaches from below $\Delta=\frac32$. When the singlet hits $\Delta=\frac32$, in the large $N_f$ approximation, the quartic $SU(N_f)$-invariant operator in bQED hits $\Delta=3$ from below: the physical interpretation is that tricritical QED merges with the $\bC\bP^{N_f-1}$ model. 

In the fermionic QED instead, it is the $SU(N_f)$-adjoint operator that approaches $\Delta=\frac32$, from above. Correspondingly, decreasing $N_f$, quartic interactions (which, crucially, transform in a non-trivial representation of $SU(N_f)$ and hence break the global symmetry) hit $\Delta=3$ from above: the physical interpretation is that  fQED merges with QED-GN$_-$. 

A simple estimate of the merging points is easy to obtain:\footnote{Both in the bosonic and fermionic QED's, it is a one-loop Feynman diagram, describing the decay of a meson into two photons, that gives a big contribution to the singlet operators, $+\frac{192}{3\pi^2 N_f}$.}
\be N_{bQED}^*\sim 2 \cdot \frac{128}{3\pi^2}\simeq 8.6 \qquad\quad \qquad N_{fQED}^*\sim 2 \cdot \frac{64}{3\pi^2}\simeq 4.3 \ee
In section \ref{SEC:fixedpoints} we give various less crude estimates of $N^*_f$ in all the four mergings, studying the actual operators that hit $\Delta=3$ at the merging point, which are quartic in the flavors or quadratic in the Hubbard-Stratonovich fields. We consistently find that in the bosonic QEDs $N_f^*\sim 9{-}11$, while in fermionic QEDs $N_f^*\sim 3{-}7$.\footnote{In any case we are not able to say anything conclusive about the fate of fermionic QED at $N_f=4$, which has many interesting physical applications in condensed matter \cite{Lee:2018udi, Song:2018ccm, Song:2018ial}.} 
 
In the fQED$-$QED-GN$_-$ merging, we use available scaling dimensions at $2^{nd}$ order in $1/N_f$ \cite{Gracey:1993sn, Gracey:2018fwq}. It turns out the $2^{nd}$ order corrections have the same sign of the $1^{st}$ order ones. This implies that going to $2^{nd}$ order increases the estimate of $N_{f}^*$. This fact, together with an analysis of the bosonic mergings in $4{-}\epsilon$ dimensions, suggests that a \emph{square-root ansatz} for the scaling dimension (enforcing the square root behavior of $\Delta$ at $N_f \to N_f^*$) might be better than a simple \emph{linear extrapolation ansatz}, see section \ref{sec:improveesti}.
  
One of the main points of the paper is that the merging pattern suggested by large $N_f$ arguments (no symmetry breaking in bosonic QEDs and symmetry breaking in fermionic QED) is in agreement with the pattern dictated by the $N_f=2$ dualities.

 \vspace{0.2cm}
 
 Let us close this discussion comparing with other large $N_f$ $2{+}1d$ models.

In $O(N)$ models or $O(N)$-Gross-Neveu models, the $1^{st}$ order corrections to the singlet operators are smaller, ${\sim}\frac{32}{3\pi^2N}$, and there is a unitary CFT for all $N \geq 1$. Yukawa and quartic scalar interactions are weaker than gauge interactions.

 In the minimally supersymmetric QED with $N_f$ flavors\footnote{The theory is $\cN{=}1$ QED with zero superpotential. The matter content consists of $N_f$ fermionic flavors, $N_f$ bosonic flavors and a gauge invariant Majorana fermion. The non-gauge interactions are cubic $SU(N_f)$-invariant Yukawa couplings.} \cite{Benve:2018A} the $1^{st}$ order correction to the $SU(N_f)$-singlet quadratic operator, instead of being large as in non supersymmetric QED's, is zero. Moreover, setting $N_f{=}2$, the $1^{st}$ order scaling dimensions of all quadratic and quartic operators agree pretty well with the scaling dimensions of a dual $\cN{=}1$ Gross-Neveu-Yukawa model \cite{Gaiotto:2018yjh, Benini:2018bhk}, computed in the $D{=}4{-}\e$ expansion \cite{Benini:2018bhk}. Since bosonic theories merge without symmetry breaking, and fermionic theories merge with symmetry breaking, a supersymmetric theory must live the single life.
 
On the other hand, it is natural to expect that non supersymmetric gauge theories with non-Abelian gauge groups, and possibly Chern-Simons interactions, display a qualitative behavior similar to QED. The large-$N_f$ expansion might be useful for instance to improve our understanding of the quantum phase scenarios of \cite{Komargodski:2017keh, Gomis:2017ixy, Choi:2018tuh}.







\newpage

\section{Fixed points of "easy-plane"-QED's, $U(N_f/2)^2$ symmetry}\label{SEC:fixedpoints}
In this section we describe our interacting bosonic and fermionic QED's with $U(N_f/2)^2$-invariant quartic couplings. The theories live in $2{+}1$ dimensions and all the flavors are massless. We report the scaling dimensions of simple scalar operators, computed at $1^{st}$ order in the large-$N_f$ expansion \cite{Benve:2018A} (some of the results were already computed in \cite{Vas:1983, Gracey:1993ka, Gracey:1993sn, Xu:2008, Kaul:2008xw, Chester:2016ref, Gracey:2018fwq}). Recent advances in the large-$N_f$ limit of QED$_3$'s include \cite{Klebanov:2011td, Pufu:2013vpa, Dyer:2015zha, Diab:2016spb, Giombi:2016fct, Chester:2017vdh}. For recent investigations of QED$_3$'s in the context of $N_f{=}2$ dualities or quantum critical points see \cite{Braun:2014wja, Janssen:2017eeu, Lee:2018udi, Ihrig:2018ojl, Zerf:2018csr, Gracey:2018fwq}. 

We first discuss the RG fixed points in the ungauged models, where the existence of four unitary fixed points can be established rigorously for any $N_f>1$. 

Upon gauging the $U(1)$ symmetry, the RG flow structure is the same for large enough $N_f$, but for small $N_f$ the fate of the gauged fixed points can be different. We estimate in each case the $N_f^*$ where the real fixed points merge into a pair of complex fixed points. The merging is driven by mesonic operators becoming relevant and entering the action.\footnote{It is conceivable that a similar mechanism is at play with monopole operators (this would break the $U(1)_{top}$ topological symmetry).  In this paper we disregard the possibility that monopoles enter the action. This is certainly the correct thing to do if the gauge group is non-compact ($\bR$ instead of $U(1)$), since in this case monopoles do not exist. Studying possible mergings driven by monopoles is an interesting project that goes beyond the scope of this paper.} When $N_f<N_f^*$, the RG flow slows down passing close to the complex conjugate pair of complex CFTs. In the case of bosonic QED, the RG flow eventually experiences a first order phase transition, with, as far as we can tell, the same global symmetry of the UV theory. In the case of fermionic QED, in the IR we have symmetry breaking, and the RG flows eventually reach the Non-Linear-Sigma-Model with target space the complex Grassmannian
\be \frac{U(N_f)}{U(N_f/2) \times U(N_f/2) } \,.\ee



\subsection{Bosonic QED}
We start with bosonic QED, with $N_f/2$ flavors $\phi_i$ plus  $N_f/2$ flavors $\phit_i$, all of gauge charge $1$, organizing all the fixed points in the plane of quartic couplings of the potential
\be \label{pot1} V= \l_1 ( (\sum|\phi_i|^2)^2 + (\sum|\tilde{\phi}_i|^2)^2) + 2 \l_2  (\sum|\phi_i|^2)(\sum|\tilde{\phi}_j|^2) \ee
This potential preserves $U(N_f/2)^2 \rtimes \bZ_2^e$ symmetry, where $\phi_i$, $\tilde{\phi}_j$ are in the fundamental of the two $SU(N_f/2)$, and $\bZ_2^e$ exchanges $\phi_i \lra \tilde{\phi}_i$. These symmetries prevent other couplings to be generated. On the locus $\l_1 = \l_2$ the global symmetry is enhanced to $U(N_f)$.

Let us first consider the ungauged model, with $2N_f$ real scalars and global symmetry is $(O(N_f) \times O(N_f)) \rtimes \bZ_2^e$, becoming $O(2N_f)$ on the locus $\l_1 = \l_2$. There are four fixed points:
\begin{enumerate}
\item{Free fixed point}, with $\l_1=\l_2=0$. Both quartic couplings are relevant, obviously.
\item{Decoupled fixed point}, with $\l_1>0, \l_2=0$. It describes two decoupled $O(N_f)$ models. We know from the numerical bootstrap \cite{Kos:2016ysd} that $\Delta[|\phi|^2_{singlet}]_{O(N_f)}>\frac32$ (if $N_f>1$\footnote{ In the $O(n)$ vector model, the rigorous scaling dimensions of the quadratic singlet operator is $1.412625(10)$ if $n=1$ (Ising model), $1.5117(25)$ for the $O(2)$-model, $1.5957(55)$ for the $O(3)$-model, \cite{Kos:2016ysd}  and goes up to $\sim 2 - \frac{32}{3 \pi^2 n}$ at large $n$. Notice the different qualitative structure at $n=1$.}), so $\Delta[\sum|\phi_i|^2 \sum|\tilde{\phi}_j|^2]_{decoupled}= 2 \Delta[|\phi|^2_{singlet}]_{O(N_f)} >3$. This proves rigorously that, for any $N_f>1$, this fixed point is attractive.
\item{$O(2N_f)$ model}, with $\l_1 = \l_2 > 0$. $O(2N_f)$ global symmetry. A relevant symmetry breaking quartic deformation, $(\sum|\phi_i|^2-|\tilde{\phi}_i|^2)^2$, drives the theory to the decoupled fixed point.
\item{"Model-3"}with $\l_1 >0,  \l_2 < 0$. Global symmetry is $(O(N_f) \times O(N_f)) \rtimes \bZ_2^e$. A relevant quartic deformation triggers an RG flow to the decoupled fixed point.
\end{enumerate}
The RG flows looks as follows
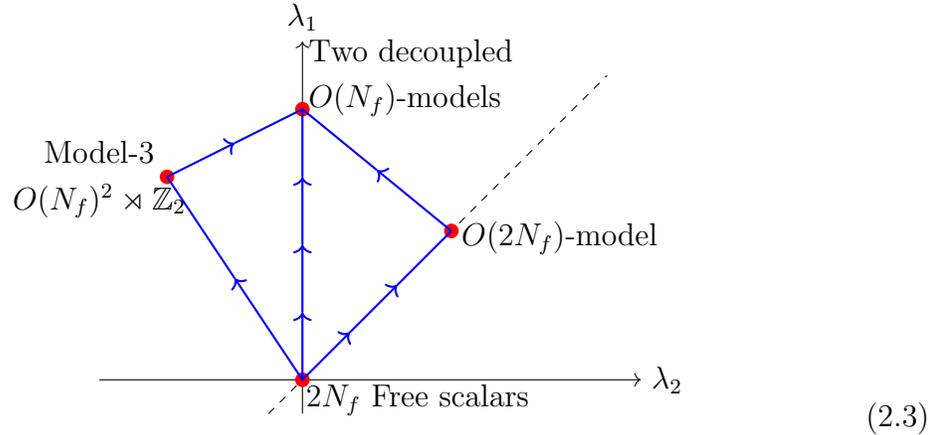
\be \label{RGON}\bpic[scale=0.9]
   \coordinate (y) at (0,5);
    \coordinate (x) at (5,0);
    \draw[->] (-3,0) --  (x) node[right] {$\l_2$};
    \draw[->]  (0,-0.5) --  (y) node[above] {$\l_1$};
    \draw[-,dashed,thin] (-0.5,-0.5) -- (4.5,4.5);
  \draw [fill, red] (0,0) circle [radius=0.1];
  \draw [fill, red] (0,4) circle [radius=0.1];
  \draw [fill, red] (2.2,2.2) circle [radius=0.1];
  \draw [fill, red] (-2,3) circle [radius=0.1];
  \draw[->, blue, thick] (0,0) -- (0,1);   \draw[->, blue, thick] (0,1) -- (0,2);  \draw[->, blue, thick] (0,2) -- (0,3);   \draw[-, blue, thick] (0,3) -- (0,4);
  \draw[->, blue, thick] (0,0) -- (0.7,0.7);  \draw[->, blue, thick] (0.7,0.7) -- (1.4,1.4);   \draw[-, blue, thick] (1.4,1.4) -- (2.2,2.2); 
  \draw[->, blue, thick] (0,0) -- (-1,1.5);  \draw[-, blue, thick] (-1,1.5) -- (-2,3); 
  \draw[->, blue, thick] (2.2,2.2) -- (1.1,3.1);  \draw[-, blue, thick] (1.1,3.1) -- (0,4); 
  \draw[->, blue, thick] (-2,3) -- (-1,3.5);  \draw[-, blue, thick] (-1,3.5) -- (0,4); 
  \node at (3.8,2.1) {$O(2N_f)$-model};
  \node at (1.7,-0.3) {$2N_f$ Free scalars};
  \node at (1.6,4.5) {$\ba{l}\text{Two decoupled} \\ O(N_f)\text{-models} \ea$};
  \node at (-3,3) {$\ba{c}\text{Model-} 3 \\ O(N_f)^2 \rtimes \bZ_2 \ea$};
  \epic\ee
Let us emphasize that this is an exact result valid for any $N_f>1$.  The pattern agrees with the findings of \cite{Calabrese:2002bm}.

\paragraph{Gauging the $U(1)$ symmetry at even $N_f$.}
When we gauge the global symmetry the four fixed points flow to four interacting QED fixed points.\footnote{In the $4-\e$ expansion, tricritical QED is described by a small $\l_1=\l_2 \sim 1/N_f^2$ fixed point, the ep-bQED has $\l_1 \sim 1/N_f$, $\l_2 \sim 1/N_f^2$, while the other two fixed points have $\l_1 \sim 1/N_f$, $\l_2 \sim 1/N_f$.} 
The global symmetry includes a topological (or magnetic) $U(1)_{top}$, under which only monopole operators are charged. The full UV global symmetry is
\be\label{exactsymmb1} \left( \frac{SU(N_f/2) \times SU(N_f/2)  \times U(1)_b \rtimes \bZ^e_2}{\bZ_{N_f}  }   \times U(1)_{top} \right) \rtimes \bZ^\cC_2 \ee
$\bZ_{N_f}$ acts as $\{\phi_i, \phit_i\} \ra e^{2\pi i/N_f} \{\phi_i, \phit_i\}$, which is a gauge transformation, so we need to quotient by this factor (it does not act on $U(1)_{top}$ because in bosonic QED the bare monopoles are gauge invariant, so the monopoles are not dressed).  $U(1)_b$: $\{\phi_i, \phit_i\} \ra  \{e^{i \a} \phi_i, e^{- i \a}\phit_i\}$. $\bZ_2^e$ ($\phi_i \leftrightarrow \tilde{\phi}_i$) does not commute with the symmetries appearing on its left. $\bZ_2^{\CC}$ is the charge-conjugation symmetry: $\phi_i \ra \phi^*_i, A_\mu \ra - A_\mu$. There is also time-reversal (or parity) symmetry $\bZ_2^\CT$.

It is convenient to rewrite the quartic potential using two Hubbard-Stratonovich real scalars $\sigma_{\pm}$:
\be  \CL =  \frac{1}{4 e^2} F_{\mu \nu} F^{\mu \nu } + \sum_{i=1}^{N_f/2} (|D_\mu \phi_i|^2+|D_\mu \tilde{\phi}_i|^2)  + \sum_\pm \sigma_\pm  \sum_{i=1}^{N_f/2}( |\phi_i|^2  \pm |\tilde{\phi}_i|^2)  - \frac{\eta_1}{2}(\s_+^2 + \s_-^2) - \eta_2 (\s_+^2 - \s_-^2)\,.  \ee
Integrating out $\sigma_\pm$, one recovers the potential \ref{pot1}, with $\{\l_1, \l_2\}$ expressed in terms of $\{\eta_1, \eta_2\}$. At large enough $N_f$ the first and the last two terms are irrelevant (their $N_f=\infty$ scaling dimension is $4$), so it is enough to work with
\be  \CL =   \sum_{i=1}^{N_f/2} (|D_\mu \phi_i|^2+|D_\mu \tilde{\phi}_i|^2)  + \sum_\pm \sigma_\pm  \sum_{i=1}^{N_f/2}( |\phi_i|^2  \pm |\tilde{\phi}_i|^2) \,.  \ee
where the photon and the Hubbard-Stratonovich fields $\s_\pm$ have an effective propagator obtained resumming a geometric series of Feynman bubble graphs.

If $N_f$ is large enough, the qualitative features of the RG flows are not changing when turning on the U(1) gauge coupling, which triggers an RG flow from \ref{RGON} to four interacting bQED's:
\be\label{bQEDRG}\bpic
  \draw [fill, red] (0,0) circle [radius=0.1];
  \draw [fill, red] (0,4) circle [radius=0.1];
  \draw [fill, red] (2.2,2.2) circle [radius=0.1];
  \draw [fill, red] (-2,3) circle [radius=0.1];
  \draw[->, blue, thick] (0,0) -- (0,1);   \draw[->, blue, thick] (0,1) -- (0,2);  \draw[->, blue, thick] (0,2) -- (0,3);   \draw[-, blue, thick] (0,3) -- (0,4);
  \draw[->, blue, thick] (0,0) -- (0.7,0.7);  \draw[->, blue, thick] (0.7,0.7) -- (1.4,1.4);   \draw[-, blue, thick] (1.4,1.4) -- (2.2,2.2); 
  \draw[->, blue, thick] (0,0) -- (-1,1.5);  \draw[-, blue, thick] (-1,1.5) -- (-2,3); 
  \draw[->, blue, thick] (2.2,2.2) -- (1.1,3.1);  \draw[-, blue, thick] (1.1,3.1) -- (0,4); 
  \draw[->, blue, thick] (-2,3) -- (-1,3.5);  \draw[-, blue, thick] (-1,3.5) -- (0,4); 
  \node at (3.8,2.1) {$\ba{l}\text{bQED}_+ \\ (\bC\bP^{N_f-1}\text{-model}) \\ U(N_f) \ea$};
  \node at (1.7,-0.3) {$\ba{l}\text{Tricritical bQED} \\ U(N_f) \ea$};
  \node at (1.4, 4.7) {$\ba{l}\text{ep-bQED}\, (\text{"easy plane" QED}) \\ U(N_f/2)^2 \ea$};
  \node at (-3.1,3) {$\ba{c}\text{bQED}_- \\ U(N_f/2)^2  \ea$};
  \epic\ee
Assuming that below a certain $N_f^*$ two or four fixed points become complex, the picture of the RG flows below $N_f^*$ is different for the RG flows between complex conjugates CFT's, but there still are RG flows from the complex conjugated pair coming from bQED --- bQED$_+$ to  the complex conjugated pair coming from bQED$_-$ --- ep-bQED.

At the fixed points bQED and bQED$_+$ the global symmetry  is enhanced to
\be\label{exactsymmb2} \left( \frac{SU(N_f)}{\bZ_{N_f}} \times U(1)_{top} \right) \rtimes \bZ_2^{\CC} \ee
where $\bZ_{N_f}$ is the center of $SU(N_f)$.
All gauge invariant local operators, including the monopoles, transform in $SU(N_f)$ representations with zero $N_f$-ality.

\subsubsection*{The two fixed points with $U(N_f)$ symmetry}
The scaling dimensions of simple scalar operators in the large-$N_f$ limit, at the fixed points with $U(N_f)$ symmetry, are \cite{Benve:2018A, Vas:1983}:
\be \label{tableBQED} \ba{|c|l|}\hline
\ba{c} \text{bQED (tricritical)} \\ U(N_f)\text{-symmetry}\ea  & 
 \ba{l}  \Delta[\phi^*\phi_{SU(N_f)-adjoint}] = 1 - \frac{64}{3 \pi^2 N_f}  \\
 \Delta[|\phi|^2_{SU(N_f)-singlet}] = 1 + \frac{128}{3 \pi^2 N_f}  \\
  \Delta[\phi^*_i\phi^*_j\phi^k \phi^l - \text{traces}] = 2 - \frac{128}{3 \pi^2 N_f}   \\
 \Delta[|\phi|^4_{SU(N_f)-singlet}] = 2 + \frac{256}{3 \pi^2 N_f}   \\
  \ea \\  \hline
\ba{c} \text{bQED}_+ (\bC\bP^{N_f-1} \text{model}) \\ U(N_f)\text{-symmetry} \ea  & 
 \ba{l}  \Delta[\phi^*\phi_{SU(N_f)-adjoint}] = 1 - \frac{48}{3 \pi^2 N_f} \\
\Delta[\phi^*_i\phi^*_j\phi^k \phi^l - \text{traces}] = 2 - \frac{48}{3 \pi^2 N_f}  \\
  \Delta[\s_+] = 2 - \frac{144}{3 \pi^2 N_f}    \\ 
  \Delta[\frac{-5 \mp \sqrt{37}}{12}\s_+^2 + F^{\mu\nu}F_{\mu\nu}] = 4 - \frac{32(4 \pm \sqrt{37})}{3 \pi^2 N_f}   \ea \\ \hline
  \ea \ee
The quartic operators $[\phi^*_i\phi^*_j\phi^k \phi^l - \text{traces}]$ transform in the "adjoint-$2$" representation of $SU(N_f)$, with Dinkyn labels $[2,0,\ldots,0,2]$.

We are not aware of any exact $2^{nd}$ order computation in bosonic QED's. Extrapolating finite-$N_f$ numerical simulations, \cite{Kaul:2011dqx} estimated the $2^{nd}$ order correction to the adjoint in bQED$_+$ to be\footnote{This is taken from $\eta_N$ in the caption of figure $5$ of \cite{Kaul:2011dqx}, where there seems to by a sign typo.}
\be \label{2ndorderNUM} \Delta[\phi^*\phi_{SU(N_f)-adj}] = 1 - \frac{48}{3 \pi^2 N_f} + \frac{1.8(2)}{N_f^2}\,.\ee

The merging of these two fixed points happens when the $|\phi|^4_{singlet}$ operator (that at $N_f=\infty$ has $\Delta=2$) in bQED, decreasing $N_f$, hits $\Delta=3$ from below, and the $\s_+^2$ operator (that at $N_f=\infty$ has $\Delta=4$) in bQED$_+$ hits $\Delta=3$ from above.  Actually, the operator $\s_+^2$ mixes strongly with $F^{\mu\nu}F_{\mu\nu}$, that also has $\Delta=4$ at $N_f=\infty$. The mixing was studied in \cite{Vas:1983}, from which we take the results in the last line of \ref{tableBQED}.

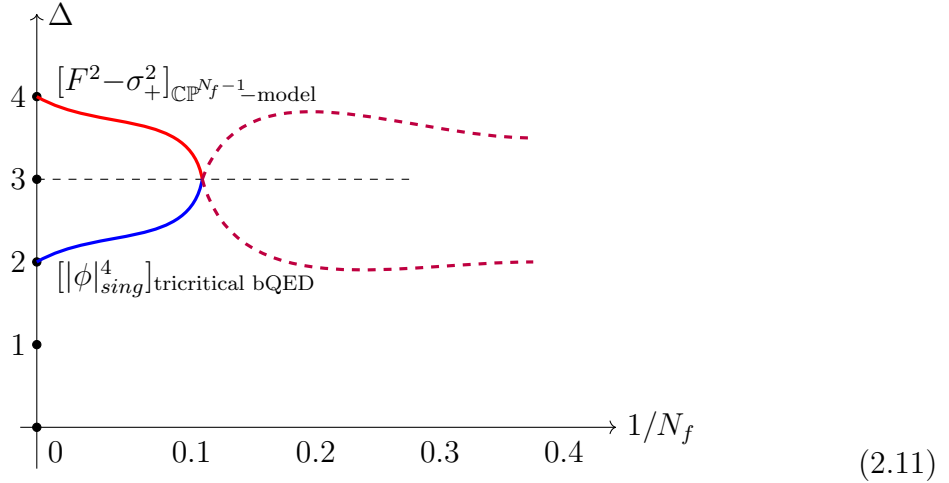
\be\bpic[scale=1.1]
   \coordinate (y) at (0,5);
    \coordinate (x) at (7,0);
    \coordinate (m) at (2,3);
    \draw[->] (-0.2,0) --  (x) node[below,right] {$1/N_f$};
    \draw[->]  (0,-0.5) --  (y) node[right] {$\Delta$};
    \draw[-,dashed,thin] (0,3) -- (4.5,3); 
  \draw [fill] (0,4) circle [radius=0.05];  \draw [fill] (0,3) circle [radius=0.05];  \draw [fill] (0,2) circle [radius=0.05];  \draw [fill] (0,1) circle [radius=0.05];  \draw [fill] (0,0) circle [radius=0.05];
  \draw[-, red, very thick] (0,4) to [out=-30,in=100] (m);
  \draw[-, blue, very thick] (0,2) to  [out=30,in=-100] (m);
  \draw[-, purple, very thick, dashed]  (m) to [out=-75,in=180] (6,2);  
  \draw[-, purple, very thick, dashed] (m) to [out=70,in=180] (6,3.5);  
  \node at (0,1) [left] {$1$};  \node at (0,2) [left] {$2$};  \node at (0,3) [left] {$3$};  \node at (0,4) [left] {$4$};
  \node at (0,-0.3) [below,right] {$0$};
  \node at (1.5,-0.3) [below,right] {$0.1$};
  \node at (3,-0.3) [below,right] {$0.2$};
  \node at (4.5,-0.3) [below,right] {$0.3$};
    \node at (6,-0.3) [below,right] {$0.4$};
    \node at (0.1,4.15) [right] {$[F^2 {-} \s_+^2]_{\bC\bP^{\!N_{\!f}{-}1}\!\!-\text{model}}$};
    \node at (0.1,1.8) [right] {$[|\phi|^4_{sing}]_{\text{tricritical bQED}}$};
  \epic\ee

Imposing that the interactions reach marginality we can estimate $N_f^*$: 
\be\label{estim1} \Delta[[|\phi|^4_{SU(N_f)-singlet}]_{bQED} = 3 \quad \ra \quad N_f^* \sim \frac{256}{3\pi^2}\simeq 8.6 \ee
\be\label{estim2}  \Delta[-0.924 \s_+^2 + F^{\mu\nu}F_{\mu\nu} ]_{bQED_+} = 3 \quad \ra \quad N_f^* \sim \frac{32(4 + \sqrt{37})}{3 \pi^2} \simeq 10.9 \ee

Another way to estimate the merging point is to impose that the scaling dimension of the singlet bilinear in bQED is equal to the scaling dimension of the Hubbard-Stratonovich field $\s_+$ in bQED$_+$:
\be \label{estim3}\Delta[|\phi|^2_{SU(N_f)-singlet}]_{bQED} = 1 + \frac{128}{3 \pi^2 N_f} =  \Delta[\s_+]_{bQED_+} = 2 - \frac{144}{3 \pi^2 N_f} \quad \ra \quad N_f^* \sim 9.2 \ee
Even if these three arguments are not completely independent, it is encouraging to get somewhat consistent results.

\subsubsection*{The two fixed points with $U(N_f/2)^2$ symmetry}
Let us now move to  the fixed points with $U(N_f/2)^2$ symmetry, the scaling dimensions of the mesonic gauge invariant operators are \cite{Benve:2018A}:\footnote{See \cite{Benve:2018A} for the anomalous dimensions of the remaining quartic operators.}
\be \label{tableBQED1} \ba{|c|l|}\hline
\ba{c} \text{bQED}_-  \\ U(N_f/2)^2\text{-symmetry} \ea  & 
 \ba{l}  \Delta[\phi^*\phi_{SU(N_f/2)-adj}, \phit^*\phit_{SU(N_f/2)-adj}] = 1 - \frac{48}{3 \pi^2 N_f}  \\  
 \Delta[\phi^*_i\phit_j, \phi_i\phit^*_j] = 1 - \frac{72}{3 \pi^2 N_f}  \\
   \Delta[\phi^*_i\phi^*_j\phit_k\phit_l, \phi_i\phi_j\phit^*_k\phit^*_l] = 2 - \frac{144}{3 \pi^2 N_f}  \\
 \Delta[\sum_{i=1}^{N_f/2}  |\phi_i|^2  +  |\tilde\phi_i|^2]   = 1 + \frac{144}{3 \pi^2 N_f}  \\ 
 \Delta[(\sum_{i=1}^{N_f/2}  |\phi_i|^2  +  |\tilde\phi_i|^2)^2]   = 2 + \frac{288}{3 \pi^2 N_f}  \\ 
   \Delta[\s_-] = 2 + \frac{48}{3 \pi^2 N_f}   \ea \\ \hline
       \ba{c}  \text{easy plane bQED}  \\ U(N_f/2)^2\text{-symmetry} \ea  & 
 \ba{l}  \Delta[\phi^*\phi_{SU(N_f/2)-adj}, \phit^*\phit_{SU(N_f/2)-adj}] = 1 - \frac{32}{3 \pi^2 N_f}  \\ 
  \Delta[\phi^*_i\phit_j, \phi_i\phit^*_j] = 1 - \frac{56}{3 \pi^2 N_f}  \\
  \Delta[\phi^*_i\phi^*_j\phit_k\phit_l, \phi_i\phi_j\phit^*_k\phit^*_l] = 2 - \frac{64}{3 \pi^2 N_f}  \\
   \Delta[\s_-] = 2 + \frac{32}{3 \pi^2 N_f}  \\ 
    \Delta[\s_+] = 2 - \frac{160}{3 \pi^2 N_f}    \ea  \\ \hline
\ea \ee
Imposing that the singlet bilinear in bQED$_-$ meets the Hubbard-Stratonovich field $\s_+$ in ep-bQED:
\be \Delta[\sum_{i=1}^{N_f/2}  |\phi_i|^2  +  |\tilde\phi_i|^2]_{bQED_-}  =  \Delta[\s_+]_{ep-bQED}  \quad \ra\quad N_f^* \sim 10.3 \ee

Unfortunately in this case we do not have scaling dimensions of the pair of operators $\{\s_-^2, F^{\mu\nu}F_{\mu\nu}\}$. From the quartic operator in bQED$_-$ hitting $\Delta=3$ from below we get
\be  \Delta[(\sum_{i=1}^{N_f/2}  |\phi_i|^2  +  |\tilde\phi_i|^2)^2]_{bQED_-} = 3 \quad \ra \quad N_f^*\sim 9.7 \ee

Let us also consider the possibility of a different merging pattern, for instance that bQED$_+$ merges with ep-bQED and breaks the global symmetry. It is easy to see that the scaling dimensions disfavour this scenario: in bQED$_+$, the anomalous dimension of $[\phi^*_i\phi^*_j\phi^k \phi^l - \text{traces}]$ is negative, so decreasing $N_f$ such operators do not hit $\Delta=3$, which would be required in order for bQED$_+$ to merge with ep-bQED.

\subsubsection{Improved estimate of $N_f^*$? A square-root ansatz}\label{sec:improveesti}
If the annihilation-of-fixed-points scenario is correct, it must be that the scaling dimensions of the various operators $\Delta[\cO](N_f)$ present a square root behaviour when $N_f \searrow N_f^*$, and the anomalous dimensions becomes complex when $N_f<N_f^*$. For instance for the quartic singlet operator in tricritical bQED, we might use a simple ansatz of the form
\be\label{ansatz} \Delta[|\phi|^4_{singlet}]_{bQED} = 3 - \sqrt{1 - N_f^* /N_f} \sim 2 + \frac{N_f^*}{2 N_f} + \frac{(N_f^*)^2}{8 N_f^2} + \frac{(N_f^*)^3}{16 N_f^3} + O(1/N_f^4)\ee

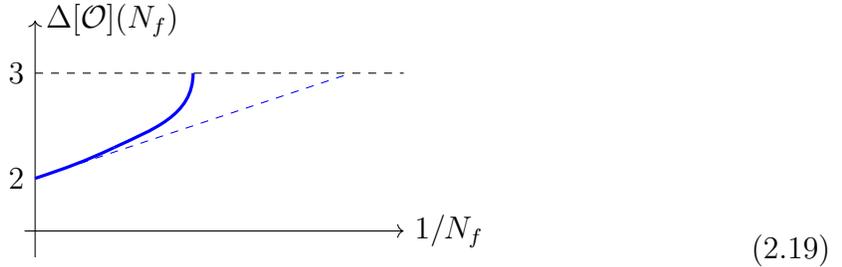
\be \bpic[scale=0.7]
   \coordinate (y) at (0,4);
    \coordinate (x) at (7,0);
    \coordinate (m) at (3,3);
    \coordinate (n) at (1.5,1.586);
    \draw[->] (-0.2,0) --  (x) node[below,right] {$1/N_f$};
    \draw[->]  (0,-0.5) --  (y) node[right] {$\Delta[\cO](N_f)$};
    \draw[-,dashed,thin] (0,3) -- (7,3); 
  \draw[-, blue, very thick] (0,1) to  [out=18.45,in=180+25.24] (n);
   \draw[-, blue, very thick] (n) to  [out=25.24,in=-90] (m);
  \draw[-, blue, dashed] (0,1) to  [out=18.45,in=180+18.45] (6,3);
   \node at (0,1) [left] {$2$};  \node at (0,3) [left] {$3$};  
  \epic\ee

Notice that this ansatz predicts that all the higher order corrections have the same sign of the $1^{st}$ order correction.

Using the $1^{st}$ order result $ \Delta[|\phi|^4_{singlet}] = 2 + \frac{256}{3 \pi^2 N_f}$, in the square-root ansatz \ref{ansatz} provides the estimate $N_f^* = 2 \cdot \frac{256}{3 \pi^2} \sim 17.3$. This is a factor of $2$ larger than the estimate 
in \ref{estim1}, which used a linear extrapolation. For all the operators in all the models considered in this section, the square-root ansatz \ref{ansatz} provides estimates of $N_f^*$ which are a factor of $2$ larger than the estimates using the linear extrapolation.

Let us emphasize that including the square root behavior at $N_f \to N_f^*$ is equivalent to imposing information about strongly coupled phenomena. It would be desirable to have  scaling dimensions at higher order in $1/N_f$: this would allow to test if  ansatze that include the square root behavior (of the form $ \Delta = f(1/N_f) - g(1/N_f) \sqrt{1 - N_f^* /N_f} $, where $f$ any $g$ are analytic functions) are better than the naive extrapolation. 

Let us add a couple of observations.

In the case of fQED and QED-GN$_-$ (see \ref{sec:fQED}),  some $2^{nd}$ order results are available: they have the same sign of the $1^{st}$ order corrections (this gives support to the square-root ansatz and the merging scenario). Accordingly, the estimate for $N_f^*$ using the $2^{nd}$ order result \ref{festim2} is $\sim 1.5$ times larger than the estimate using the $1^{st}$ order result.

In the case of the Abelian Higgs model, in the $4-\epsilon$ expansion it is known that the zeroes of the one loop beta function of the quartic coupling $\lambda|\phi|^4$ are given by
\be \frac{\lambda_*}{2 \pi ^2 } = \frac{N_f+18 \pm \sqrt{ N_f^2-180 N_f-540}}{N_f (N_f+4)} \epsilon\,. \ee
The "$+$" solution is the bQED$_+$, the "$-$" solution is the tricritical bQED. From the previous equation it follows that in the limit $\epsilon \to 0^+$, the exact result for the fixed point merging is $N_f^*= 6 (15+4 \sqrt{15}) \sim 183$. On the other hand, we can perform a computation analogous to eqs. \ref{estim1}, \ref{estim2}, \ref{estim3}, in $d \to 4^-$, using the generic-$d$ scaling dimensions computed in \cite{Vas:1983}. The result is $N_f^*(d \to 4^-) \sim 90$, which is indeed a factor of $\sim 2$ smaller than the exact result. This computation tells us that, in dimension $d \to 4^-$, the square-root ansatz \ref{ansatz} is better than the linear extrapolation, suggesting that the same might be true in dimension $3$, and the linear extrapolation underestimates $N_f^*$ also in $d=3$.

\subsubsection*{Singlet sextic interactions of bosonic tricritical points} At the tricritical fixed point the sextic  $SU(N_f)$-singlet operator at infinite $N_f$ has $\Delta=3$. The $1^{st}$ order correction is 
\be \Delta[(\sum_{i=1}^{N_f/2}  (\phi^i \phi^*_i +  \tilde\phi^i \tilde\phi^*_i))^3] = 3 + 3 \frac{128}{3\pi^2N_f} + O(1/N_f^2) \ee
So the sextic $SU(N_f)$ invariant deformation is irrelevant. Modulo tuning mass and quartic term to zero, tricritical bQED is a \emph{stable fixed point}. At the merging of the tricritical fixed point with the critical fixed point sextic singlet interactions do not play a role.
 $3d$ bosonic gauge theories at the tricritical point (with quartic interactions tuned to zero) were studied in a completely different regime in \cite{Aharony:2018pjn, Dey:2018ykx}, where they named the model \emph{regular boson theory}. \cite{Aharony:2018pjn, Dey:2018ykx} found that for $U(N_c)_k$ Chern-Simons with $1$ bosonic flavor, at large $N_c$ and large $k$ with $N_c/k$ fixed, there is a stable fixed point and possibly (depending on the value of $N_c/k$) an unstable fixed point. Combining these two  results, it is natural to suggest that at finite $N_c, N_f, k$, bosonic QCD always has a stable \emph{tricritical}, or \emph{regular}, fixed point. 

\subsection{Fermionic QED}\label{sec:fQED}
We consider fermionic QED with $N_f/2$ flavors $\psi_i$ plus  $N_f/2$ flavors $\psit_i$ (each $\psi, \psit$ is a complex two-component  $3d$ fermion). The quartic Gross-Neveu interactions are modeled by Yukawa cubic couplings with two real Hubbard-Stratonovich scalar fields, $\rho_+$ and $\rho_-$.\footnote{\label{Fclari} Much of the existing literature considers QEDs with $N$ four-component Dirac fermions $\Psi_i$, $i=1,\ldots, N$, in generic dimension $d$. In $d=3$, the global symmetry can be $U(2N)$ or $U(N)^2$, depending on the precise form of the Yukawa (or Gross-Neveu-Yukawa) couplings. 

In terms of two-component $3d$ fermions $\Psi_i = (\psi_i, \psit_i)$ and $\bar{\Psi}_i = (\bar{\psi}_i, - \bar{\psit}_i)$. So $\sum_{i=1}^{N} \bar{\Psi}_i \Psi_i= \sum_{i=1}^{N}  (\bar\psi^i \psi_i -  \bar{\tilde\psi}^i \tilde\psi_i)$ is a $U(N)$-singlet in $d \neq 3$, but it is part of the $SU(2N)$-adjoint in $d=3$. 

On the other hand $\sum_{i=1}^{N} \bar{\Psi}_i \Gamma_5 \Psi_i= \sum_{i=1}^{N}  (\bar\psi^i \psi_i + \bar{\tilde\psi}^i \tilde\psi_i)$ is a $SU(2N)$-singlet in $d=3$.

Often, what is called QED-Gross-Neveu has $\cL_{int}= \sigma \sum_{i=1}^{N} \bar{\Psi}_i \Psi_i$, with $U(N)^2$ global symmetry in $d=3$. We instead named this model QED-GN$_-$. On the other hand \cite{Wang:2017txt} calls QED-Gross-Neveu the model that we named QED-GN$_+$, with $d=3$ global symmetry $U(2N)$.}  $\rho_+$ and $\rho_-$ are parity-odd, and all our theories are parity invariant. The Lagrangian reads
\be \CL = \frac{1}{4 e^2} F_{\mu \nu} F^{\mu \nu } + \sum_{i=1}^{N_f/2} (\bar{\psi}^i \slashed{D} \psi_i+\bar{\psit}^i \slashed{D} \psit_i) +\sum_\pm \r_\pm  \sum_{i=1}^{N_f/2} (\bar{\psi}^i \psi_i \pm \bar{\psit}^i\tilde{\psi}_i ) + \ldots \ee
The $\ldots$ stand for quartic interactions and kinetic terms for the $\r_\pm$ fields. The mass terms for $\r_\pm$ are relevant at large enough $N_f$.

We start discussing the ungauged model, with $O(N_f)^2 \rtimes \bZ^e_2$ global symmetry, the RG flows between the $4$ fixed points are triggered by mass terms for the scalars $\r_\pm$.

There are $4$ fixed points, similar to the bosonic case: a free theory, a decoupled fixed point with both $\r_+$ and $\r_-$ (renaming $\r_{\pm}= \r \pm \tilde{\r}$, it splits into two decoupled $O(N_f)$-invariant Gross-Neveu models), a Gross-Neveu fixed point with only $\r_-$ and $O(N_f)^2 \rtimes \bZ_2^e$-symmetry, and a Gross-Neveu fixed point with only $\r_+$ and $O(2N_f)$-symmetry.

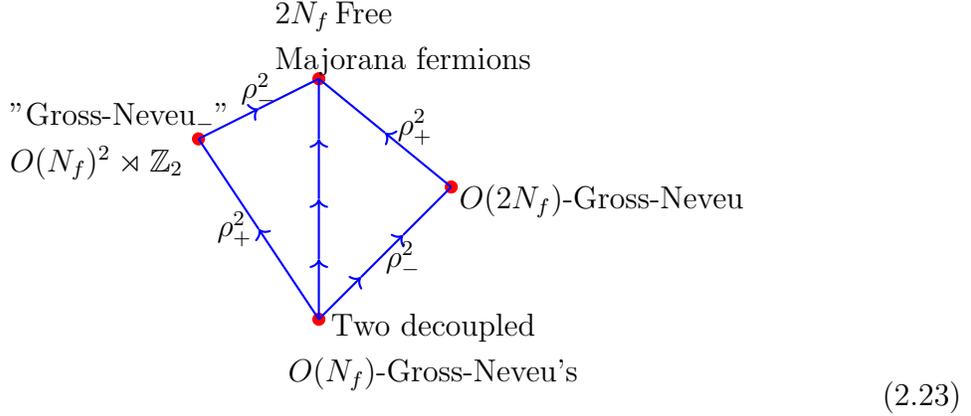
\be\label{GNRG}\bpic[scale=0.8]
  \draw [fill, red] (0,0) circle [radius=0.1];
  \draw [fill, red] (0,4) circle [radius=0.1];
  \draw [fill, red] (2.2,2.2) circle [radius=0.1];
  \draw [fill, red] (-2,3) circle [radius=0.1];
  \draw[->, blue, thick] (0,0) -- (0,1);   \draw[->, blue, thick] (0,1) -- (0,2);  \draw[->, blue, thick] (0,2) -- (0,3);   \draw[-, blue, thick] (0,3) -- (0,4);
  \draw[->, blue, thick] (0,0) -- (0.7,0.7);  \draw[->, blue, thick] (0.7,0.7) -- (1.4,1.4);   \draw[-, blue, thick] (1.4,1.4) -- (2.2,2.2); 
  \draw[->, blue, thick] (0,0) -- (-1,1.5);  \draw[-, blue, thick] (-1,1.5) -- (-2,3); 
  \draw[->, blue, thick] (2.2,2.2) -- (1.1,3.1);  \draw[-, blue, thick] (1.1,3.1) -- (0,4); 
  \draw[->, blue, thick] (-2,3) -- (-1,3.5);  \draw[-, blue, thick] (-1,3.5) -- (0,4); 
  \node at (4.7,2) {$\ba{l} O(2N_f)\text{-Gross-Neveu} \ea$};
  \node at (-3.3,3) {$\ba{l}\text{"Gross-Neveu$_-$"} \\ O(N_f)^2 \rtimes \bZ_2 \ea$};
  \node at (1.4, 4.7) {$\ba{l}2N_f\, \text{Free} \\ \text{Majorana fermions} \ea$};
  \node at (1.9,-0.5) {$\ba{c} \text{Two decoupled} \\ O(N_f)\text{-Gross-Neveu's}  \ea$};  
  \node at (-1.4,1.5) {$\r_+^2$};
  \node at (1.6,3.2) {$\r_+^2$};
  \node at (1.4,1) {$\r_-^2$};
  \node at (-1,3.8) {$\r_-^2$};
  \epic\ee

For any $N \geq 1$, it is known with good accuracy that in the $O(N)$ Gross-Neveu model, $\Delta[\r^2]<3$ (at large $N$ $\Delta[\r^2] \sim 2+\frac{16}{3\pi^2N}$, at $N=1$ $\Delta[\r^2] \sim 1.59$), so in particular the deformations $\r_+^2$ and $\r_-^2$ are relevant.

\paragraph{Gauging the $U(1)$ symmetry at even $N_f$.}
The global symmetry becomes
\be\label{exactsymmf1} \frac{\left(SU(N_f/2) \times SU(N_f/2) \times U(1)_b \times \uot \right) \rtimes \bZ_2^e}{\bZ_{N_f}} \rtimes \bZ_2^{\CC} \,,\ee
where $\bZ_{N_f}$ is a gauge transformation acting as 
\be\label{ZNF} \{\psi_i, \tilde{\psi}_i, \M_{bare} \} \ra  \{ e^{2 \pi i /N_f} \psi_i, e^{2 \pi i /N_f} \tilde{\psi}_i , - \M_{bare} \}\ee
 The quotient acts also on $U(1)_{top}$ because bare monopole operators $\M_{bare}$ are not gauge invariant. The gauge invariant monopoles with minimal topological charge, $\M^{\pm 1}$, are dressed with $N_f/2$ fermionic zero-modes and are invariant under \ref{ZNF} \cite{Cordova:2017kue}. $U(1)_b$: $\{\psi_i, \psit_i\} \ra  \{e^{i \a} \psi_i, e^{- i \a}\psit_i\}$. $\bZ_2^e$: $\psi_i \leftrightarrow \tilde{\psi}_i$. $\bZ_2^{\CC}$ is the charge-conjugation symmetry.  There is also a time-reversal (or parity) symmetry that satisfies a non-trivial algebra, see \cite{Cordova:2017kue} for pure QED, adding Yukawa interactions does not change their results.  

As in the bosonic case, if $N_f$ is large enough, gauging the $U(1)$ symmetry triggers an RG flow from \ref{GNRG} to the following $4$ interacting fermionic fixed points:
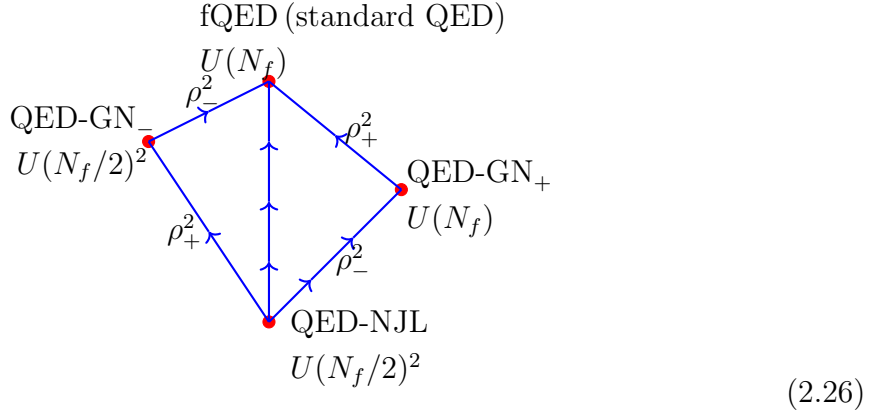
\be\label{fQEDRG}\bpic[scale=0.8]
  \draw [fill, red] (0,0) circle [radius=0.1];
  \draw [fill, red] (0,4) circle [radius=0.1];
  \draw [fill, red] (2.2,2.2) circle [radius=0.1];
  \draw [fill, red] (-2,3) circle [radius=0.1];
  \draw[->, blue, thick] (0,0) -- (0,1);   \draw[->, blue, thick] (0,1) -- (0,2);  \draw[->, blue, thick] (0,2) -- (0,3);   \draw[-, blue, thick] (0,3) -- (0,4);
  \draw[->, blue, thick] (0,0) -- (0.7,0.7);  \draw[->, blue, thick] (0.7,0.7) -- (1.4,1.4);   \draw[-, blue, thick] (1.4,1.4) -- (2.2,2.2); 
  \draw[->, blue, thick] (0,0) -- (-1,1.5);  \draw[-, blue, thick] (-1,1.5) -- (-2,3); 
  \draw[->, blue, thick] (2.2,2.2) -- (1.1,3.1);  \draw[-, blue, thick] (1.1,3.1) -- (0,4); 
  \draw[->, blue, thick] (-2,3) -- (-1,3.5);  \draw[-, blue, thick] (-1,3.5) -- (0,4); 
  \node at (3.5,2.1) {$\ba{l}\text{QED-GN}_+  \\ U(N_f) \ea$};
  \node at (1.5,-0.4) {$\ba{l}\text{QED-NJL} \\ U(N_f/2)^2  \ea$};
  \node at (1.4, 4.7) {$\ba{l}\text{fQED} \, (\text{standard QED}) \\ U(N_f) \ea$};
  \node at (-3.1,3) {$\ba{c}\text{QED-GN}_-  \\ U(N_f/2)^2  \ea$};  
  \node at (-1.4,1.5) {$\r_+^2$};
  \node at (1.55,3.2) {$\r_+^2$};
  \node at (1.4,1) {$\r_-^2$};
  \node at (-1.1,3.8) {$\r_-^2$};
  \epic\ee
The global symmetry at the fixed points fQED and QED-GN$_+$ is enhanced to
\be\label{exactsymmf2} \frac{SU(N_f) \times \uot}{\bZ_{N_f}} \rtimes \bZ_2^{\CC} \,.\ee

\subsubsection*{Fermionic QED and its partner}
The scaling dimensions of mesonic scalar gauge invariant operators to leading order in the large-$N_f$ limit are \cite{Gracey:1993sn,Xu:2008, Chester:2016ref, Gracey:2018fwq, Benve:2018A}:\footnote{Let us observe that, at $1^{st}$ order in $1/N_f$, the anomalous (not the total) dimensions of the fermionic fixed points \ref{tableFQED}, \ref{tableFQED1} are equal to the anomalous dimensions of the bosonic fixed points \ref{tableBQED}, \ref{tableBQED1}. This is true for the Hubbard-Stratonovich fields and for quadratic operators in the charged fields.}
\be \label{tableFQED} \ba{|c|l|}\hline
\ba{c} \text{fQED} \\ U(N_f) \ea &
 \ba{l}  \Delta[\bar\psi \psi_{SU(N_f)-adj}] = 2 - \frac{64}{3 \pi^2 N_f} + \frac{256(28-3\pi^2)}{9 \pi^4 N_f^2} \sim 2-\frac{2.16}{N_f}-\frac{0.47}{N_f^2} \\ 
 \Delta[\bar\psi \psi_{SU(N_f)-singlet}]  = 2 + \frac{128}{3 \pi^2 N_f} \\
 \Delta[|\psi|^4_{[0,1,0\ldots,0,1,0]}] = 4- \frac{192}{3\pi^2N_f} \\
 \Delta[|\psi|^4_{[2,0,\ldots,0,2]}] = 4 + \frac{64}{3\pi^2N_f}  \\
 \Delta[\{(|\psi|^2_{singlet})^2 , F^{\mu\nu}F_{\mu\nu}\}] = 4 + \frac{64(2\pm \sqrt{7})}{3\pi^2N_f} \ea \\ \hline
\ba{c} \text{QED-GN}_- \\ U(N_f/2)^2 \ea & 
      \ba{l} \Delta[\bar\psi \psi_{SU(N_f/2)-adj}] = 2 - \frac{48}{3 \pi^2 N_f} + \frac{64(100-9\pi^2)}{9 \pi^4 N_f^2} \sim 2 -\frac{1.62}{N_f}+\frac{0.82}{N_f^2}   \\ 
      \Delta[\bar\psi^i \tilde\psi_j, \bar{\tilde\psi}^i  \psi_j] = 2 - \frac{72}{3 \pi^2 N_f}  \\
      \Delta[\sum_{i=1}^{N_f/2}  (\bar\psi^i \psi_i +  \bar{\tilde\psi}^i \tilde\psi_i)] = 2 + \frac{144}{3 \pi^2 N_f}  \\
      \Delta[\r_-] = 1 + \frac{48}{3 \pi^2 N_f}  - \frac{8(1232-243\pi^2)}{9 \pi^4 N_f^2} \sim 1 + \frac{1.62}{N_f} + \frac{10.64}{N_f^2}\\
      \Delta[\r_-^2] = 2 + \frac{144}{3 \pi^2 N_f} \ea   \\ \hline
 \ea \ee
The quartic fermionic operators in fQED were computed in \cite{Xu:2008, Chester:2016ref}, we indicated the Dinkyn labels of the $SU(N_f)$ representation under which they transform. The mixing between the quartic singlet and $F^{\mu\nu}F_{\mu\nu}$ is strong also here, and was solved in \cite{Chester:2016ref}, the lowest eigenvalue of the singlets does not seem to run fast enough to hit $\Delta=3$ (which would suggest the symmetry-preserving merging fQED$-$QED-GN$_+$). We also included the order $1/N_f^2$ contributions, when known \cite{Gracey:1993sn, Gracey:2018fwq}.\footnote{\cite{Gracey:1993sn} studies pure QED and eq. $27$ gives the scaling dimension of $\sum_{i=1}^{N} \bar{\Psi}_i \Psi_i= \sum_{i=1}^{N}  (\bar\psi^i \psi_i -  \bar{\tilde\psi}^i \tilde\psi_i)$ which is part of the $SU(N_f{=}2N)$-adjoint in $d=3$. See footnote \ref{Fclari}.

The QED-GN$_-$ results are given eqs. $4.4$ and $4.6$ of \cite{Gracey:2018fwq}, which studies a model (referred to as QED-Gross-Neveu in \cite{Gracey:2018fwq}) with $\cL_{int}= \sigma \sum_{i=1}^{N} \bar{\Psi}_i \Psi_i$. When $d=3$ this model is what we call QED-GN$_-$, with $U(N)^2$ $3d$ global symmetry. So the results of \cite{Gracey:2018fwq} are valid for our QED-GN$_-$ with $N_f{=}2N$ flavors. Moreover, \cite{Gracey:2018fwq} reports the dimension of $\sum_{i=1}^{N} \bar{\Psi}_i \Psi_i$,  an operator which vanishes on-shell because of the equation of motion of $\sigma$. We report the scaling dimension of $\sigma$, using the relation $\Delta[\s]=3-\Delta[\sum_{i=1}^{N} \bar{\Psi}_i \Psi_i]$.} The $2^{nd}$ order corrections to the adjoint are quite small, while $\r_-$ receives a big contribution at $2^{nd}$ order, from which \cite{Gusynin:2016som, Kotikov:2019rww} estimated chiral symmetry breaking below $N_f^* \sim 2.8469\cdot 2= 5.69$ in fQED.


The conjectural merging of the two fixed points fQED and QED-GN$_-$ happens when, decreasing $N_f$, the lowest quartic fermion operator ($|\psi|^4_{[0,1,0,\ldots,0,1,0]}$) hits $\Delta=3$ from above and the mass term of the Hubbard-Stratonovich field $\r_-^2$ hits $\Delta=3$ from below: 
\be \Delta[|\psi|^4_{[0,1,0\ldots,0,1,0]}] _{fQED} = 3 \quad \ra \quad N_f^* \sim \frac{192}{3\pi^2} \simeq 6.5 \ee
\be  \Delta[\r_-^2]_{QED-GN_-} = 3 \quad \ra \quad N_f^*\sim \frac{144}{3\pi^2}\simeq 4.9 \ee

Another estimate of the point of merging comes equating the adjoint in fQED with $\r_-$ in QED-GN$_-$, using the $2^{nd}$ order anomalous dimensions we get
\be \label{festim2}\Delta[\bar\psi \psi_{SU(N_f)-adj}]_{fQED} = \Delta[\r_-]_{QED-GN_-}  \quad \ra \quad N_f^* = \frac{56+2\sqrt{678 \pi ^2-3472}}{3 \pi ^2} \sim 5.72 \ee
Had we used the $1^{st}$ order anomalous dimensions, we would have got $N_f^* \sim 3.8$. Hence, the $2^{nd}$ order in $1/N_f$ corrections in fermionic QEDs increase the value of the merging point. This is because the $2^{nd}$ order corrections have the same sign of the $1^{st}$ order corrections, both in $\Delta[\bar\psi \psi_{SU(N_f)-adj}]_{fQED}$ and in  $\Delta[\r_-]_{QED-GN_-}$ (this fact gives more evidence to the square root behaviour which must be present if the merging scenario is correct, as discussed in section \ref{sec:improveesti}).

Studying fermionic QED at finite $N_f$ but continuos dimension $d$, \cite{Herbut:2016ide} estimated $N_f^*\sim 2.89 \cdot 2 = 5.8$, while \cite{Giombi:2015haa} found an upper bound for the merging: $N_f^* < 4.4 \cdot 2= 8.8$.

\subsubsection*{QED-GN$_+$ and its partner}
We now move to the last fixed points. The scaling dimensions for QED-GN$_+$ and QED-NJL are \cite{Benve:2018A}:
 \be \label{tableFQED1} \ba{|c|l|}\hline
\ba{c} \text{QED-GN}_+ \\ U(N_f) \ea &
 \ba{l}  \Delta[\bar\psi \psi_{SU(N_f)-adj}]  = 2 - \frac{48}{3 \pi^2 N_f}  \\  \Delta[\r_+] = 1 - \frac{144}{3 \pi^2 N_f} \\  \Delta[\r_+^2] = 2 - \frac{240}{3 \pi^2 N_f} \ea   \\ \hline
\ba{c} \text{QED-NJL} \\ U(N_f/2)^2 \ea  &
 \ba{l}  \Delta[\bar\psi \psi_{SU(N_f/2)-adj}] = 2 - \frac{32}{3 \pi^2 N_f}  \\ \Delta[\bar\psi^i \tilde\psi_j, \bar{\tilde\psi}^i  \psi_j] = 2 - \frac{56}{3 \pi^2 N_f}  \\
   \Delta[\r_-] = 1+ \frac{32}{3 \pi^2 N_f}  \\  \Delta[\r_+] = 1 - \frac{160}{3 \pi^2 N_f} \\
    \Delta[\r_+ \r_-] =  2 -  \frac{32}{3 \pi^2 N_f}    \\
 \Delta[\rho_+^2+(4 \mp \sqrt{17})\rho_-^2] =  2 -  \frac{16(5 \pm 3\sqrt{17})}{3 \pi^2 N_f} 
 \ea   \\ \hline
\ea \ee
We can estimate $N_f^*$ in two ways. First, imposing that the adjoint in QED-GN$_+$ meets the singlet in QED-NJL:
\be \Delta[\bar\psi \psi_{SU(N_f)-adj}]  = 2 - \frac{48}{3 \pi^2 N_f} =  \Delta[\r_-] = 1+ \frac{32}{3 \pi^2 N_f} \quad \ra \quad N_f^*  \sim 2.7 \ee
It is conceivable that, as in \ref{festim2}, including $2^{nd}$ order anomalous dimensions moves this estimate up significantly. Second, looking at when $\r_-^2$ (after having solved the mixing with $\r_+^2$) hits $\Delta=3$ from below:
\be  \Delta[\r_-^2 + 0.123 \r_+^2]_{QED-NJL} =  2 + \frac{16(3\sqrt{17}-5)}{3 \pi^2 N_f} = 3 \quad \ra \quad N_f^* \sim \frac{16(3\sqrt{17}-5)}{3 \pi^2} \simeq 4 \ee

\section{Boson $\lra$ fermion dualities for QED$_3$ with $2$ flavors}\label{SEC:dualities}
\cite{Karch:2016sxi, Wang:2017txt} discovered that $3d$ QED's with two fermionic or two bosonic flavors satisfy two different  boson $\lra$ fermion dualities. We start from the duality between $N_f{=}2$ QED-GN$_+$ and bosonic QED$_+$ ($\bC\bP^1$ model) and from this duality we obtain the other three dualities. The dualities are a conjecture, and require the gauge groups to be compact ($U(1)$ instead of $\bR$).

Even if $N_f=2$ is very small, we will try to compare large $N_f$ scaling dimensions for mesonic and monopole operators with numerical results and expectations from dualities. It turns out that the monopole scaling dimensions agree surprisingly well with expectations from duality and with numerical simulations. As for the mesons, the operators transforming in the adjoint seem to behave relatively well, consistently with the fact that the adjoint receives small corrections from the $N_f=\infty$ value. On the other hand, operators which are singlets of the global symmetry group do not agree with expectations from duality and numerical simulations. This is related to the fact that both $1^{st}$ and $2^{nd}$ order corrections in $1/N_f$ seems to be large for singlets.

Since the fixed points are expected to be complex, the scaling dimensions should be complex. But large-$N_f$ perturbative computations naively only provide real anomalous dimensions. Using Pad\'e-style resummations, with branch cuts in the ansatz like the ones we discussed in section \ref{sec:improveesti}, it might be possible to circumvent this issue. In this section we content ourselves with real scaling dimensions, which should be interpreted as the real part of the full scaling dimensions.

The large $N_f$ monopole scaling dimensions are available only for bQED$_+$ \cite{Murthy:1989ps, Dyer:2015zha} and fQED  \cite{Borokhov:2002ib, Pufu:2013vpa}, it would be very interesting to study monopoles also in the other six models considered in this paper, they might provide many new constraints on the values of $N_f^*$.



\subsection{QED-GN$_+$ $\lra$ bosonic QED ($\bC\bP^1$ model)}
The duality we start from was discussed in detail in \cite{Wang:2017txt}. On the fermionic side the model is called QED-Gross-Neveu. On the bosonic side the model is also called "non-compact" $\bC\bP^1$ model.
\be \label{dual2} \ba{c} U(1) \, + \, 2\, \psi's \\ V= \rho_+ (\bar{\psi}_1\psi_1+ \bar{\psi}_2\psi_2) \ea \quad  \Longleftrightarrow \quad \ba{c} U(1) \, +  \, 2\, \phi's \\ V = \sigma_+ (|\phi_1|^2 + |\phi_2|^2 ) \ea \ee
We denote the fermionic flavors $\psi_1, \psi_2$ and the bosonic flavors $\phi_1, \phi_2$.\footnote{Integrating out the Hubbard-Stratonovich real scalar $\s_+$ we would get $V = (|\phi_1|^2 + |\phi_2|^2 )^2$, and the second mapping in \ref{map2} would be $ \bar{\psi}_1\psi_1- \bar{\psi}_2\psi_2  \Llra  |\phi_1|^2 + |\phi_2|^2 $.}
The mapping of the simplest operators, which is soon going to be very useful for us, is
\be \label{map2} \left\{\ba{c} \rho_+ \\ \bar{\psi}_1\psi_1- \bar{\psi}_2\psi_2 \ea \right\} 
\Llra 
 \left\{\ba{c} |\phi_1|^2 - |\phi_2|^2 \\  \sigma_+  \ea \right\}\ee
This map \cite{Wang:2017txt} follows from the structure of massive deformations of the fixed point, or deriving the duality from the basic bosonization duality.
The global symmetries $\frac{SU(2) \times U(1)_{top}}{\bZ_2}\rtimes \bZ_2^\CC$ and $(\frac{SU(2)}{\bZ_2} \times U(1)_{top}) \rtimes \bZ_2^\CC$ enhance at the fixed point to $SO(5)$ \cite{Wang:2017txt}.

It is possible to argue for the enhanced $SO(5)$ symmetry using a self-duality of the $\bC\bP^1$ model, self-duality that follows from the old particle $\lra$ vortex duality \cite{Nahum:2015vka}.

The $4$ monopoles of QED-GN$_+$ combine with $\r_+$ to form a ${\bf 5}$ of $SO(5)$ and map to the monopoles plus the $3$ adjoint mesons in bQED$_+$:\footnote{To be more precise, $\M$ is a complex operator, with $\M^\dagger = \M^{-1}$, and the operators appearing in the ${\bf 5}$, which is a real irrep of $SO(5)$, are $\Re[\M]$ and $\Im[\M]$. Same comment for $\M_\psi, \phi_1^*\phi_2, \phi_2^*\phi_1, \ldots$.}
 \be \{\M^{+1}_{\psi_1}, \M^{+1}_{\psi_2} , \M^{-1}_{\bar{\psi}_1}, \M^{-1}_{\bar{\psi}_2}, \r_+\} \quad \Llra \quad \{\M^{+1}, \M^{-1}, \phi_1^*\phi_2, \phi_2^*\phi_1, |\phi_1|^2 - |\phi_2|^2 \} \ee
$\M^q$ denotes the monopole of topological charge $q$. The monopole scaling dimension in large-$N_f$ bosonic QED$_+$ \cite{Murthy:1989ps, Dyer:2015zha} (the $2^{nd}$ order in $1/N_f$ \cite{Dyer:2015zha} result reproduces well the lattice results of \cite{LSK, Kaul:2011dqx} at small $N_f$) give $\Delta[\M^{\pm 1}]= 0.125 \cdot 2 + 0.382 \sim 0.63$. The large-$N_f$ results for the mesonic operators instead give $\Delta[\r_+] \sim 1-\frac{144}{3\pi^2 2}$ (\ref{tableFQED1}), and $\Delta[\phi^*\phi_{spin-1}] \sim 1-\frac{48}{3\pi^2 2}+\frac{1.8}{2^2}\sim 0.64$ (\ref{2ndorderNUM}). It is encouraging that adding the $2^{nd}$-order correction \ref{2ndorderNUM} provides a seemingly correct value for the scaling dimension of $\phi^*\phi_{spin-1}$.
 
Higher degree operators organize into the symmetric traceless, the ${\bf 14}$, of $SO(5)$:
\be \label{MAP2}
 \left\{ \M^{\pm 2}_{\psi\psi},  \, \M^{\pm 1}_\psi \r_+, \, \pbp_{spin-1},\, \r_{+}^2    \right\} 
            \Longleftrightarrow
   \left\{  \M^{\pm 2}, \, \M^\pm \phi^*\phi_{spin-1} ,\,  (\phi^*\phi)^2_{spin-2},\, \sigma_{+}   \right\}   \ee
   

The large $N_f$ anomalous dimensions of spin-$0$ operators $\r_+, \r_+^2, \s_+$, e.g. $\Delta[\r_+^2] \sim 2 - \frac{240}{3 \pi^2 2}$ (\ref{tableFQED1}), $\Delta[\s_+] \sim 2- \frac{144}{3\pi^2 2}$ (\ref{tableBQED}), are unphysical when $N_f=2$. 

On the other hand, both in the ${\bf 5}$ and in the ${\bf 14}$, the monopoles \cite{Dyer:2015zha}, the spin-$1$ and spin-$2$ operators \ref{tableFQED1}, \ref{tableBQED} agree pretty well with the expectations of the duality: 
\be \Delta[\M^{\pm 1}] \sim 0.63 \qquad \Delta[\phi^*\phi_{spin-1}]  \sim 0.64\ee
\be \Delta[\M^{\pm 2}] \sim  0.311\cdot 2 + 0.875 \sim 1.5 \qquad \Delta[\pbp_{spin-1}] \sim \Delta[(\phi^*\phi)^2_{spin-2}] \sim 2-\frac{48}{3\pi^2 2} \sim 1.2 \,.\ee

\subsection{fermionic QED $\lra$  easy plane $\bC\bP^1$ model}
We now obtain other dualities starting from the above one. The strategy is to introduce scalar fields, couple these scalar fields to the simple operators\footnote{This procedure is known under the name of "flipping" in the supersymmetric literature. It was recently applied  to minimally supersymmetric $3d$ theories in \cite{Benini:2018umh, Gaiotto:2018yjh, Benini:2018bhk}.} appearing in \ref{basicmap} and flow to new theories on both sides of the duality.

Let us introduce a scalar field $\s_-$ and couple it to the first line in \ref{map2}:
\be \delta V = \rho_+  \s_- \quad \Llra \quad \delta V = \sigma_- (|\phi_1|^2 - |\phi_2|^2)\ee
On the left hand side both $\rho_+$ and $\sigma_-$ become massive and can be integrated out. We get a new duality:
\be \label{basicdual2} \qquad\ba{c} U(1) \, + \, 2\, \psi's \\ V=0 \ea \qquad  \Longleftrightarrow \quad \ba{c} U(1) \, +  2\, \phi's \\ V = \sum_\pm \sigma_\pm (|\phi_1|^2 \pm |\phi_2|^2 ) \ea \ee
\be \label{basicmap} \left\{\ba{c} \bar{\psi}_1\psi_1+ \bar{\psi}_2\psi_2\\ \bar{\psi}_1\psi_1- \bar{\psi}_2\psi_2 \ea \right\} 
\Llra 
 \left\{\ba{c} \sigma_- \\  \sigma_+  \ea \right\}\qquad \ee
The duality relates pure $N_f{=}2$ fermionic QED with $N_f{=}2$ bosonic QED with an "easy plane" potential. This process involves only one field in the ${\bf 5}$ of $SO(5)$, so it breaks the global $SO(5)$ symmetry of the fixed point to $O(4)$. 

The charge-$\pm1$ monopoles of fQED transform in the vector of $O(4)$ and map to monopoles and "bifundamental" mesons in ep-bQED:
 \be\label{monopMAP1} \{\M^{+1}_{\psi_1}, \M^{+1}_{\psi_2} , \M^{-1}_{\bar{\psi}_1}, \M^{-1}_{\bar{\psi}_2}\} \quad \Llra \quad \{\M^{+1}, \M^{-1}, \phi_1^*\phi_2, \phi_2^*\phi_1\} \ee
 The fQED monopoles $\M^{+1}_{\psi_\a} , \M^{-1}_{\bar{\psi}_\a}$ are dressed by a charged fermion.  Let us emphasize that, on the fQED side,  the four monopoles $\M_\psi^{\pm 1}$ are degenerate even if chiral symmetry breaking $U(2) \ra U(1)^2$ occurs, because the discrete symmetries $\bZ_2^e$ ($\psi_1 \lra \psi_2$) and $\bZ_2^\CC$ ($\psi_\a \lra \bar{\psi}_\a$) are unbroken.


Large-$N_f$ monopole scaling dimension in fQED \cite{Borokhov:2002ib, Pufu:2013vpa} give $\Delta[\M^{\pm 1}_\psi]= 0.265 \cdot 2 - 0.0383 \sim 0.5$. \ref{tableBQED1} gives $\Delta[\phi^*_1\phi_2, \phi_1\phi^*_2] \sim 1 - \frac{56}{3 \pi^2 2} \sim 0.05$.

Higher degree operators are organized in the symmetric-traceless of $O(4)$:\footnote{From the symmetrized product of the two ${\bf 4}$'s we need to remove the trace $\sim |\M|^2+ |\phi_1|^2|\phi_2|^2$.}
\be\{\M^{\pm2}_{\psi \psi}, \bar{\psi}\psi_{spin-1}\} \Llra \{ \M^{\pm 2}, \M^{\pm 1} \phi_1 \phi_2^*, \M^{\pm 1} \phi_2 \phi_1^*, (\phi_2 \phi^*_1)^2, (\phi_1 \phi^*_2)^2, \s_+ \}\ee
Also in this case we get qualitative agreement in the non-singlet sector: large-$N_f$ monopoles give \cite{Pufu:2013vpa}  $\Delta[\M^{\pm 2}_{\psi\psi}]= 0.673\cdot 2 - 0.194 {\sim} 1.1$. Large-$N_f$ mesons give $\Delta[\pbp_{spin-1}] \sim 2-\frac{2.16}{2}-\frac{0.47}{2^2} \sim 0.8$ (\ref{tableFQED}) and $\Delta[(\phi_2 \phi^*_1)^2, (\phi_1 \phi^*_2)^2] \sim 2-\frac{64}{3 \pi^2 2} \sim 0.9$. Lattice computations in fQED \cite{Karthik:2016ppr} give $\Delta[\pbp_{spin-1}] = 1 \pm 0.2$. The singlet operator scaling dimension $\Delta[\s_+] \sim 2 - \frac{160}{3 \pi^2 2} \sim -0.7$ (\ref{tableBQED1}) is instead unphysical.

\subsection{QED-GN$_-$ $\lra$ bosonic QED$_-$}
Applying the same logic, we can propose two new dualities. 

From duality \ref{basicdual2} we  introduce a scalar field $\rho_-$ and couple it to the second line in \ref{basicmap}:
\be \delta V = \rho_- (\bar{\psi}_1\psi_1 - \bar{\psi}_2\psi_2) \Llra  \delta V = \rho_-\sigma_+ \ee
getting 
\be \label{dual3} \ba{c} U(1) \, + \, 2\, \psi's \\ V= \rho_- (\bar{\psi}_1\psi_1 - \bar{\psi}_2\psi_2) \ea \quad  \Longleftrightarrow \quad \ba{c} U(1) \, +  2\, \phi's \\ V = \sigma_- (|\phi_1|^2 - |\phi_2|^2 ) \ea \ee
with mapping
\be \label{map3} \left\{\ba{c} \bar{\psi}_1\psi_1 +  \bar{\psi}_2\psi_2 \\ \rho_- \ea \right\} 
\Llra 
 \left\{\ba{c} \sigma_- \\ |\phi_1|^2 + |\phi_2|^2  \ea \right\}\ee
In this case both UV global symmetries are only $\sim U(1)^2$, but the four monopoles in QED-GN$_-$ must still be degenerate, because of the discrete symmetries $\bZ_2^e \times \bZ_2^\CC$.
 \be \{\M^{+1}_{\psi_1}, \M^{+1}_{\psi_2} , \M^{-1}_{\bar{\psi}_1}, \M^{-1}_{\bar{\psi}_2}\} \quad \Llra \quad \{\M^{+1}, \M^{-1}, \phi_1^*\phi_2, \phi_2^*\phi_1\} \ee
Duality implies symmetry enhancement to $O(4)$. Higher degree operators are organized in the symmetric-traceless of $O(4)$:
\be\{\M^{\pm2}_{\psi \psi}, \bar{\psi}_1\psi_2, \bar{\psi}_2\psi_1, \r_-\} \Llra \{ \M^{\pm 2}, \M^{\pm 1} \phi_1 \phi_2^*, \M^{\pm 1} \phi_2 \phi_1^*, (\phi_2 \phi^*_1)^2, (\phi_1 \phi^*_2)^2, |\phi_1|^2 + |\phi_2|^2 \}\ee
The large-$N_f$ scaling dimensions are
$\Delta[\bar{\psi}_1\psi_2, \bar{\psi}_2\psi_1]{\sim}2 - \frac{72}{3 \pi^2 2}\sim0.8$ (\ref{tableFQED}), 
$\Delta[(\phi_2 \phi^*_1)^2, (\phi_1 \phi^*_2)^2]{\sim}2 - \frac{144}{3 \pi^2 2}\sim-0.4$ (\ref{tableBQED1}).

\subsection{QED-NJL $\lra$ tricritical bosonic QED}
The fourth and last duality  relates what bQED with QED-Nambu-Jona-Lasinio. We start from \ref{dual2} and introduce $\rho_-$ and couple it to the second line in \ref{map2}:
\be \delta V = \rho_- (\bar{\psi}_1\psi_1 - \bar{\psi}_2\psi_2) \Llra  \delta V = \rho_- \sigma_+ \,.\ee
Notice that in this way we are not breaking the UV $SU(2)$ symmetry on the bosonic side. On the r.h.s. both $\r_-$ and $\s_+$ becomes massive, and we flow to a new duality:
\be \label{dual4} \ba{c} U(1) \, + \, 2\, \psi's \\ V= \sum_{\pm }\rho_\pm (\bar{\psi}_1\psi_1 \pm \bar{\psi}_2\psi_2) \ea \quad  \Longleftrightarrow \quad \ba{c} U(1) \, +  2\, \phi's \\ V=0 \ea \ee
The mapping of the mass operators is
\be \label{map4} \left\{\ba{c} \rho_+ \\ \rho_- \ea \right\} 
\Llra 
 \left\{\ba{c} |\phi_1|^2 - |\phi_2|^2  \\ |\phi_1|^2 + |\phi_2|^2  \ea \right\}\ee
On the l.h.s., because of the $\bZ_2$ symmetries, the $4$ monopole operators are still degenerate, and they map as in duality  \ref{dual2}:
 \be \{\M^{+1}_{\psi_1}, \M^{+1}_{\psi_2} , \M^{-1}_{\bar{\psi}_1}, \M^{-1}_{\bar{\psi}_2}, \r_+\} \quad \Llra \quad \{\M^{+1}, \M^{-1}, \phi_1^*\phi_2, \phi_2^*\phi_1, |\phi_1|^2 - |\phi_2|^2 \} \ee
We conclude that the CFT enjoies $SO(5)$ global symmetry, the above operators forming the $5$-dimensional representation of $SO(5)$. Higher order operators organize into the symmetric traceless, the ${\bf 14}$, of $SO(5)$:
\be \label{MAP2}
 \left\{ \M^{\pm 2}_{\psi\psi}, \, \M^{\pm 1}_\psi \r_+, \,\bar{\psi}_1\psi_2, \bar{\psi}_2\psi_1, \r_-,\, \r_{+}^2    \right\} 
            \Longleftrightarrow
   \left\{  \M^{\pm 2}, \, \M^\pm \phi^*\phi_{spin-1} ,\,  (\phi^*\phi)^2_{spin-2},\, |\phi_1|^2 + |\phi_2|^2   \right\}   \ee
The large-$N_f$ scaling dimensions are
$\Delta[\bar{\psi}_1\psi_2, \bar{\psi}_2\psi_1]\sim 2 - \frac{56}{3 \pi^2 2} \sim 1$, (\ref{tableFQED}), 
$\Delta[(\phi^*\phi)^2_{spin-2}]\sim 2-\frac{128}{3\pi^2 2} \sim -0.2$ (\ref{tableBQED}).

\acknowledgments{We are grateful to Francesco Benini, Pasquale Calabrese and Silviu Pufu for useful discussions, and to Diego Rodriguez-Gomez for comments on the draft. S.~B. is indebted with Andrea Guerrieri for an old collaboration on related topics. This work is supported in part by the MIUR-SIR grant RBSI1471GJ ``Quantum Field Theories at Strong Coupling: Exact Computations and Applications". S.B. is partly supported by the INFN Research Projects GAST and ST$\&$FI. }

\bibliographystyle{ytphys}

\end{document}